\documentclass[prb,reprint,showpacs,amsmath,amssymb,superscriptaddress,citeautoscript]{revtex4-1}

\usepackage{graphicx}

\usepackage{xcolor}

\usepackage[colorlinks,bookmarks=false,citecolor=blue,linkcolor=red,urlcolor=blue]{hyperref}
\usepackage{multirow}
\usepackage{ulem}

\emergencystretch=\maxdimen
\begin{document}
\title{Quantum Monte Carlo study of an anharmonic Holstein model}

\author{G. Paleari}
\affiliation{Universit\'e C\^ote d'Azur, CNRS, INPHYNI, France}
\affiliation{Dipartimento di Fisica, Universit\`a degli Studi di
  Milano, via Celoria 16, 20133 Milano, Italy} 
\author{F. H\'ebert}
\email[Corresponding author: ]{frederic.hebert@inphyni.cnrs.fr}
\affiliation{Universit\'e C\^ote d'Azur, CNRS, INPHYNI, France}
\author{B. Cohen-Stead}
\affiliation{Department of Physics, University of California, Davis,
  California 95616, USA}  
\author{K. Barros}
\affiliation{ Theoretical Division and CNLS, Los Alamos National Laboratory, Los Alamos, New Mexico, 87545, USA}  
\author{R$\,$T. Scalettar}
\affiliation{Department of Physics, University of California, Davis,
  California 95616, USA}  
\author{G.$\,$G. Batrouni}
\affiliation{Universit\'e C\^ote d'Azur, CNRS, INPHYNI, France}
\affiliation{Department of Physics, National University of Singapore,
  2 Science Drive 3, 117542 Singapore} 
\affiliation{Centre for Quantum Technologies, National University of
  Singapore; 2 Science Drive 3 Singapore 117542} 
\affiliation{Beijing Computational Science Research Center, Beijing
  100193, China}

\begin{abstract}
We study the effects of anharmonicity on the physics of the Holstein
model, which describes the coupling of itinerant fermions and
localized quantum phonons, by introducing a
quartic term in the phonon potential energy.  We find that the
presence of this anharmonic term reduces the extent of the charge
density wave phase (CDW) at half-filling as well as the transition
temperature to this phase.  Doping away from half-filling, we observe
a first order phase transition between the CDW and a
homogeneous phase which is also present in the
harmonic model. In addition, we study the evolution of the
superconducting susceptibility in the doped region and show that
anharmonicity can enhance the superconducting response.
\end{abstract}

\pacs{
71.10.Hf, 
71.30.+h, 
71.45.Lr, 	
63.20.-e 
}

\maketitle

\section{Introduction}

Electron-phonon interactions in solids drive a number of quantum many
body effects.  One is conventional superconductivity
(SC)\cite{Marsiglio01,Cooper56}.  Another is the formation of
insulating charge density wave (CDW) phases
\cite{Gruner94,Monceau12,Gruner88}.  Complex Hamiltonians which
describe both many electronic orbitals and multiple phonon bands, are
typically needed to describe these phenomena in real materials.
Fortunately, simplified models can often capture the key qualitative
consequences of the electron-phonon coupling, while being much more
analytically and computationally tractable.

The Holstein Hamiltonian\cite{Holstein59} is one such model.  It
describes a single electronic band and dispersionless quantum phonons
coupled locally to the fermion density.  A considerable body of
computational work exists for the Holstein model.  Studies of the
dilute limit reveal how individual electrons are dressed by phonons,
and the effective mass and transport properties of these `polarons'
have been evaluated
\cite{kornilovitch98,kornilovitch99,alexandrov00,hohenadler04,ku02,spencer05,macridin04,romero99,bonca99}.
At higher densities, the emergence of SC at generic fillings, and
gapped CDW phases at commensurate occupations has been investigated
\cite{Peierls79, Hirsch82,
  Fradkin83,Scalettar89,Marsiglio90,Freericks93,Ohgoe17,Hohenadler19}.

The solution of even this relatively simple model is not, however,
computationally easy.  Only relatively recently have the critical
temperatures for the CDW transition been evaluated for the
square\cite{Weber2018} and cubic\cite{cohenstead20} lattices via
quantum Monte Carlo (QMC).  Likewise, the determination of the
critical interaction strength at the quantum critical point for the
CDW transition on a honeycomb lattice is a rather new
development\cite{Zhang19}.  Analytic approaches, especially
Migdal-Eliashberg theory\cite{migdal58,eliashberg60} have been
critical to the understanding of the Holstein
Hamiltonian\cite{alexandrov01,bauer11}.  Their comparison with QMC has
been an especially useful line of investigation, especially in efforts
to determine the largest possible SC transition
temperature\cite{esterlis18}.

In the course of these studies, it has become apparent that the
Holstein Hamiltonian has a significant deficiency in some parameter
regimes.  Specifically, it has been shown \cite{Adolphs2013} that the
values of the phonon displacement reached in CDW phases could be quite
large, even reaching values comparable to the intersite spacing in the
system. Thus, the harmonic description of phononic excitations in the
medium provided by the Holstein model may not be sufficient, and the
effects of anharmonic terms on the phases of Holstein systems should
be taken into account
\cite{Chatterjee2004,Adolphs2013,Li2015-1,Li2015-2, Dee2020,Sentef2017,Sous2020,
  Uma2017,Freericks96}.  Several approaches to include anharmonic
effects have been considered, for example nonlinear coupling terms
between fermions and phonons
\cite{Adolphs2013,Li2015-1,Li2015-2,Dee2020,Sentef2017,Sous2020}, or quartic
\cite{hirsch93,Chatterjee2004,Freericks96} or Gaussian \cite{Uma2017}
contributions to the phonon potential energy.  Anharmonicity has also
been considered in the context of Migdal-Eliashberg
theory\cite{hui74,kavakozov78,mahan93}.

In infinite dimension, using a technique similar to dynamical mean
field theory (DMFT), Freericks, Jarrell, and Mahan \cite{Freericks96}
studied the effects of a simple anharmonic term in the form of an
additional quartic potential energy for the phonons.  They concluded
that a CDW phase exists for a large range of densities at low
anharmonicity, but that the CDW is gradually replaced at low and high
densities by a SC phase as the anharmonicity increases. The
half-filled system always remains in a CDW state. They also observed a
decrease of the critical temperatures at which CDW and SC phases
appear with increasing anharmonicity. Similar models have been studied
in one dimension \cite{Chatterjee2004}.

The goal of this article is to study the effects of such an additional
quartic anharmonic term on the behavior of the Holstein model in two
dimensions using a recently introduced Langevin algorithm
\cite{Batrouni19,Batrouni85}.  Unlike DMFT, the Langevin approach
handles spatial correlations in finite dimensions without introducing
systematic error.  In section III, we will introduce the Holstein model
and its anharmonic extension, as well as the methods we will use to
study the system and characterize the different phases. Section II
will be devoted to the study of the behavior at half-filling,
especially the CDW phase and how it evolves with
anharmonicity. Section IV will concentrate on the behavior away from
half-filling, discussing possible CDW phases as well as
superconducting behavior. We will then give some final thoughts and
conclusions.

\section{Model and methods}

We study a generalized version of the Holstein Hamiltonian which
incorporates anharmonicity in a specific way, namely as an additional
term in the quantum phonon potential energy\cite{Freericks96}:
\begin{eqnarray}
H-\mu N &=&-t \sum_{\langle ij\rangle\sigma}
\left(c^\dagger_{i\sigma}c^{\phantom{\dagger}}_{j\sigma}+
h.c.\right)-\mu \sum_{i\sigma}
n_{i\sigma}
\label{eq:ham2}\\
&&+\sum_i \left(\frac{m\omega^2 x_i^2}{2} +
\frac{p_i^2}{2m}\right) + \omega_4  \sum_i x_i^4
\label{eq:ham1}\\
&&+
\lambda \sum_{i\sigma} x_i n_{i\sigma}\label{eq:ham3}
\end{eqnarray}
The sums run over the $L^2$ sites of a two-dimensional square lattice.
The operator $c^{\phantom\dagger}_{i\sigma}$ ($c^\dagger_{i\sigma}$)
destroys (creates) a fermion of spin $\sigma =\,\uparrow {\rm or}
\downarrow$ on site $i$; $n_{i\sigma} = c_{i\sigma}^\dagger
c^{\phantom\dagger}_{i\sigma}$ is the corresponding number operator;
$x_i$ and $p_i$ are the canonical displacement and momentum operators
of the phonon mode at site $i$.  The first term (Eq.~\ref{eq:ham2})
represents the hopping energy of the fermions between neighboring
sites $\langle ij\rangle$.  A chemical potential term is included as
our algorithm performs the simulations in the grand canonical
ensemble.  The hopping parameter $t$ will be used as the energy scale.
The second term (Eq.~\ref{eq:ham1}) represents the energy of the
phonons of harmonic frequency $\omega$ and includes an anharmonic term
proportional to $x_i^4$ with a prefactor $\omega_4$, and we put $m=1$
in the rest of the article.  The third term (Eq.~\ref{eq:ham3}) is the
phonon-electron interaction.  This coupling can be rewritten as $g
\sum_{i\sigma} (a^\dagger_i + a_i) n_{i\sigma}$ where
$g=\lambda/\sqrt{2\omega}$ and $a_i$ and $a^\dagger_i$ are the
destruction and creation operators of phonons at site $i$.  We focused
on the cases where $g=1$, $\omega=0.5$ and $\omega=1$. Using two
values of $\omega$ yields the evolution of the anharmonic effects as a
function of $\omega$ and also allows comparison with previous studies
\cite{Freericks96,bradley20}.

The average value of $x_i$ on a doubly occupied site can be roughly
estimated as $- 2\lambda/\omega^2$ (see
Appendix~\ref{app:chem}).  With this expression, the ratio $\eta$ of
the anharmonic to harmonic terms is given by
\begin{equation}
\eta\equiv
\frac{16 \omega_4 \, g^2}{\omega^5} 
\label{eq:eta}
\end{equation}
For $g=1$ and $\omega<1$, $\eta$ becomes substantial even for
relatively small values of $\omega_4$.  Indeed, we will see that
$\omega_4 \lesssim 0.01$ is sufficient to affect profoundly the CDW
physics at $\omega=0.5$.

We study this model using a recently developed quantum Monte Carlo
algorithm \cite{Batrouni19} based on a Langevin equation
approach\cite{Batrouni85}. This method does not suffer from the sign
problem for the Holstein model and the scaling of the simulation time
with the number of sites is more advantageous than with conventional
methods such as determinant quantum Monte Carlo (DQMC)\cite{dqmc}. For
the Langevin algorithm applied in two dimensions, the simulation time
scales approximately as $L^{2.2}$ instead of $L^6$ for DQMC
\cite{Batrouni19}. Throughout this work, we used sizes and inverse
temperatures ranging up to $L=16$ and $\beta=20$, although, as will be
seen, it is difficult to obtain reliable results for some quantities
on large systems, especially away from half-filling or
for small values of both $\omega$ and $\omega_4$.

The Langevin approach requires a discretization of the inverse temperature $\beta$.
We used an imaginary time step $\Delta\beta=0.1$ which we checked was
sufficient so that systematic effects are smaller than statistical
error bars.  The Langevin time step was generally $dt=10^{-3}$ and we
used up to a few million Langevin steps for equilibration before performing
measurements over up to $10^7$ steps, using a standard binning of
the data to analyze statistical errors \cite{Gubernatis2016}.

We will look at the density $\rho=\sum_i\langle
n_{i\sigma}\rangle/L^2$ and its behavior as a function of $\mu$ to
detect the presence of charge gaps.  We will also examine other simple
diagonal quantities such as the average value of the phonon
displacement $\langle x_i\rangle$ and the double occupancy $\langle
n_{i\uparrow}n_{i\downarrow}\rangle$. In the harmonic case, the
particle-hole symmetry yields an analytical expression for the
chemical potential at half-filling $\mu=-\lambda^2/\omega^2$ and for
the average value of the displacement $\langle x_i\rangle = -\lambda /
\omega^2$ (see Appendix \ref{app:chem}). With $\omega_4 \ne 0$, there
is no particle-hole symmetry and the value of $\mu$ for which the
system is at half filling as well as the average displacement are
unknown and must be determined by simulations, although some rough
estimations can be made (App. \ref{app:chem}).

 \begin{figure}
 \includegraphics[width=8.5cm]{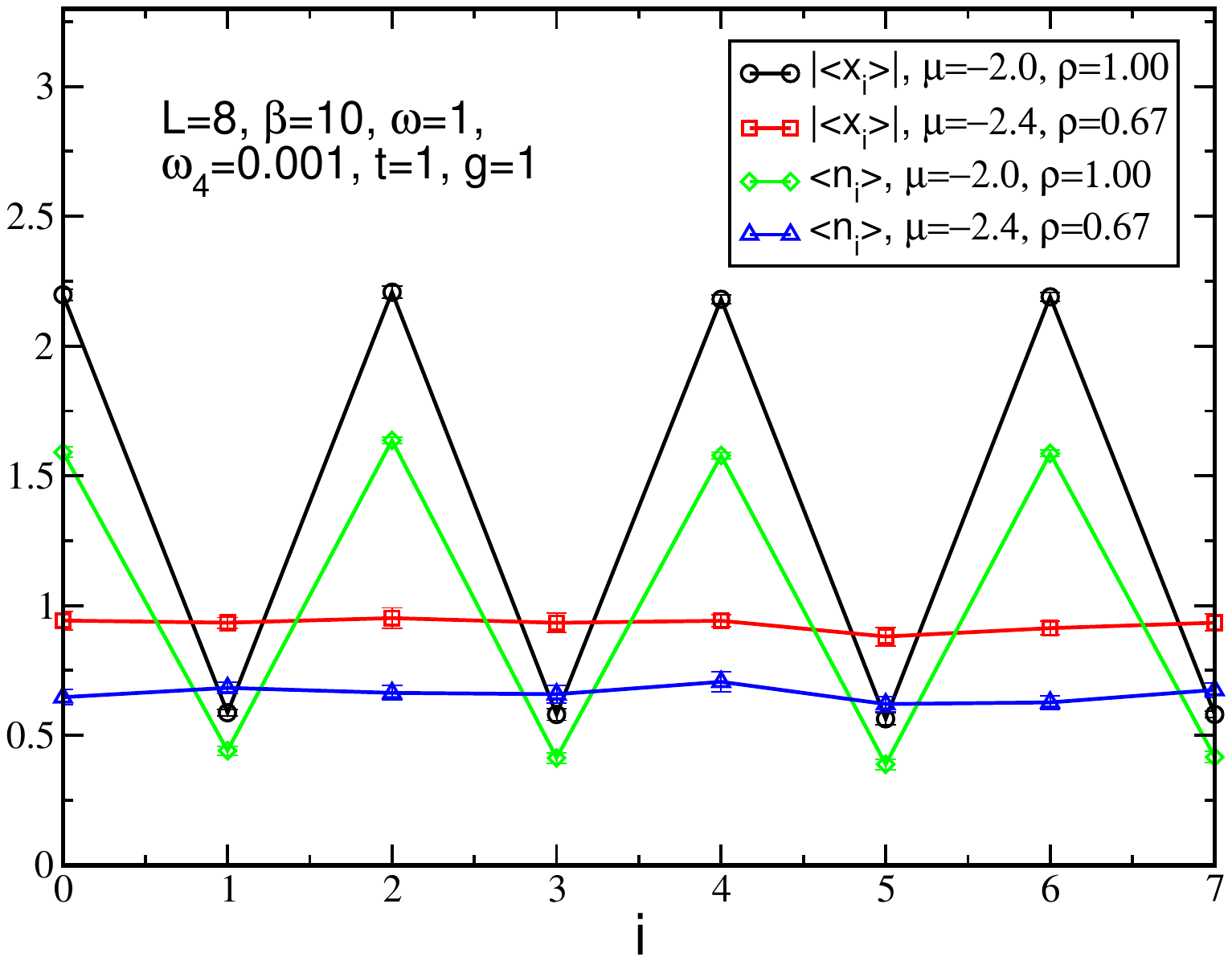}
 \caption{Behavior of the average density, $\langle n_i \rangle$, and phonon displacement,
 $|\langle x_i\rangle|$, as functions of the position, $i$, along one axis in the square lattice in the homogeneous
 and CDW phases. In CDW phase at half-filling, there is symmetry breaking and two alternate values of $\langle n_i\rangle$ and $\langle x_i\rangle$ are observed. Out of half-filling, in an homogeneous phase, $\langle n_i\rangle$ and
 $\langle x_i\rangle$ are independent of the position.\label{fig:alternate}}
 \end{figure}

To characterize the presence of a CDW phase we study the charge
structure factor, the Fourier transform at momentum $(\pi,\pi)$ of the
density-density correlation function,
\begin{equation}
S_{\rm cdw} = \sum_i\langle  n_i n_{i+j}\rangle (-1)^j \,\,.
\end{equation}
Here $n_i$ is the total number of particles on site $i$, $n_i =
n_{i\uparrow} + n_{i\downarrow}$.  The ordering vector for a
half-filled square lattice is known to be at $(\pi,\pi)$.
Incommensurate order at $q \neq (\pi,\pi)$ is possible upon doping,
but we do not see evidence of it here.  As the fermions
  enter the CDW phase, the electron-phonon coupling induces a
  corresponding order in the average values of $\langle x_i\rangle$
  where $|\langle x_i\rangle|$ takes alternatively small and large
  values on neighboring sites, following the alternating values of the
  density $n_i$ (see Fig. \ref{fig:alternate}).

Away from half-filling, the system is suspected to be superconducting,
with Cooper pairing driven by the phonons that generate on-site
attraction $U_{\rm eff}$ between particles, as noted in the discussion
of Eq.~\ref{eq:ueffandx0}.  We will look at this behavior through the
s-wave pairing susceptibility
\begin{align}
\chi_{\rm s} &= \frac{1}{L^2}\int_0^\beta d\tau \left\langle
\Delta(\tau) \Delta^\dagger(0) + {\rm h.c.}\right\rangle,
\nonumber
\\
\Delta(\tau) = \sum_i &c_{i\downarrow}(\tau) c_{i\uparrow}(\tau)
\hskip0.30in
 c_{i\sigma}(\tau) = e^{\tau H}c_{i\sigma}e^{-\tau H} \,\,.
\end{align}

\section{half-filling}

\begin{figure}
\centerline{\includegraphics[width=8.5cm]{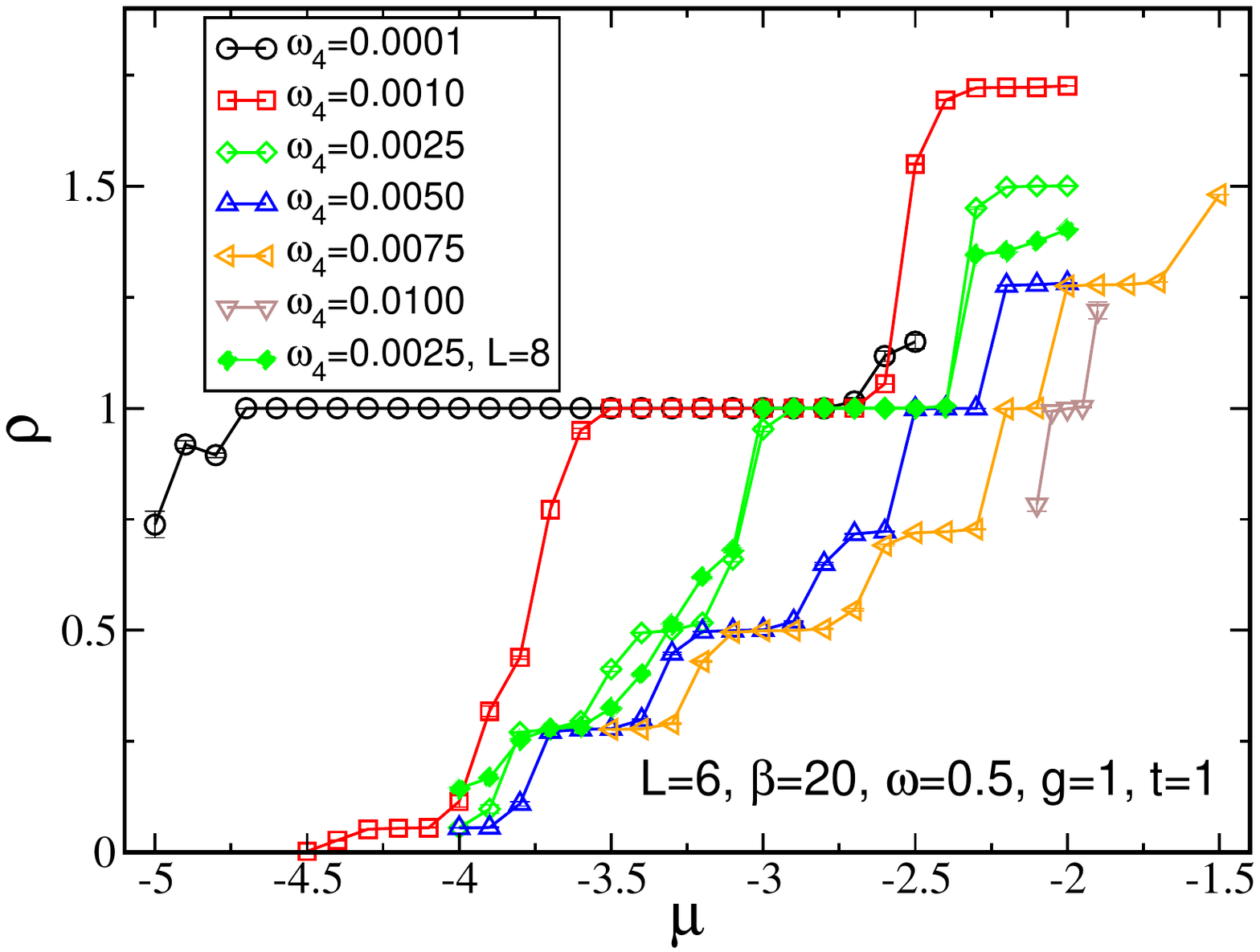}}
\caption{\label{fig:gaps}(Color online). Density as a function of
  $\mu$ for different values of $\omega_4$. $L=6$, $g=1$, $\beta=20$,
  $\omega=0.5$. We find a reduction of the charge density gap found at
  half-filling when $\omega_4$ is increased. The apparent gaps away
  from half filling are shell effects. A simulation at $L=8$,
  $\omega_4=0.0025$ shows that half-filled plateau is not affected by
  finite size effects while the plateaux off of half-filling are
  reduced for larger sizes.}
\end{figure}

Without anharmonicity, the Holstein model develops a Peierls CDW phase
at half-filling, where the chemical potential at half-filling is given
by $\mu=-2g^2/\omega$. In the presence of the anharmonic term,
however, we do not have an analytic expression for $\mu$ at half
filling (App. \ref{app:chem}).

\begin{figure}[b]
\centerline{\includegraphics[width=8.0cm]{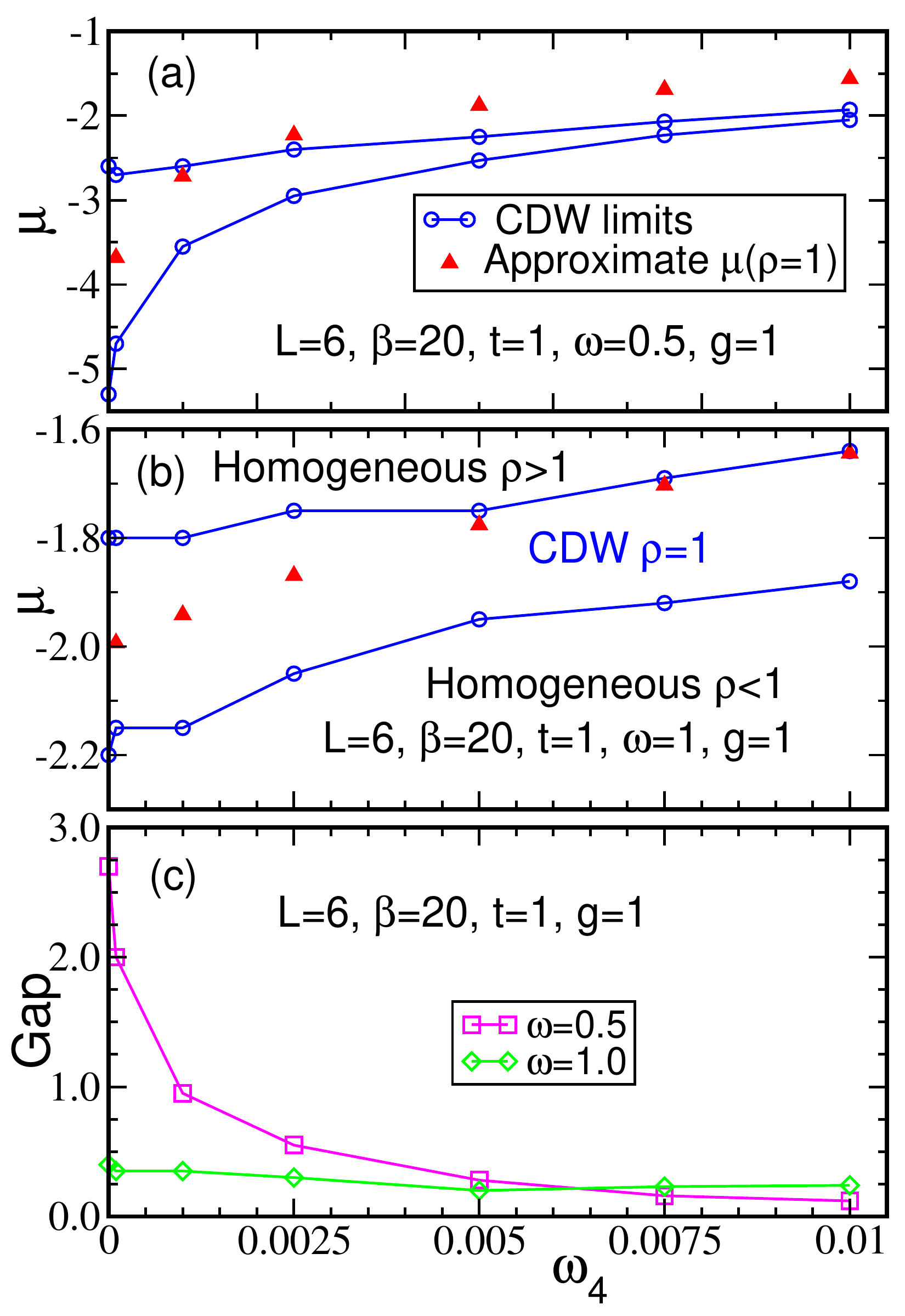}}
\caption{(Color online). Phase diagrams of the system obtained from
  $L=6$ simulations at $\beta=20$ and $g=1$, for $\omega=0.5$ (a) and
  $\omega=1$ (b) and comparison of the charge gaps in these cases (c). The anharmonic parameter $\omega_4$ ranging
  from 0 to 0.01.  The area enclosed by blue curves is the
  incompressible CDW phase at half-filling while the rest of the phase
  diagram corresponds to compressible phases that should become
  superconducting at low temperatures.  The red triangles mark the
  half-filled chemical potential inferred from the approximate theory
  of Appendix~\ref{app:chem}.  The width of the CDW phase is strongly
  reduced due to the anharmonic effects for $\omega=0.5$ (a).  For
  $\omega=1$ (b), the charge gap is relatively unaffected by the
  anharmonicity $\omega_4$ in the range shown although  it is generally smaller for $\omega=1$ than for
  $\omega=0.5$ (c).
\label{fig:phasediagram_w05}}
\end{figure}

We first study the effects of the anharmonic term (Eq.~\ref{eq:ham3})
on this phase.  To this end, we examine the evolution of the density
as a function of $\mu$ at inverse temperature $\beta=20$ which we
verified had converged to the low temperature limit. A
  large system is not needed to obtain reliable measurements of the
  charge gap at half-filling, so we used $L=6$. These simulations
also determine the value of $\mu$ for which the system is at
half-filling.  We observe (Fig.~\ref{fig:gaps}) that $\omega_4$ shifts
the insulating plateau to larger values of $\mu$ and that the width of
the half-filled density plateaux decrease with $\omega_4$. The other
smaller plateaux that are observed away from half-filling are finite
size (`shell') effects due to the finite system size.  These shell
effects are revealed by nonzero $\omega_4$ as it inhibits the CDW
order. In Fig.~\ref{fig:gaps}, results from a $L=8$ simulation for
$\omega_4=0.0025$ show that these shell effects are reduced for larger
sizes while the gap at half-filling remains essentially
unchanged. This confirms that the small plateaux appearing off of
half-filling are finite size effects, while the gap at half-filling is
not.

The plateau at half-filling is a genuine collective effect, since
there is no gap at half-filling at $g=0$.  Indeed, the
  density of states diverges there for a square lattice and only half
  of the states present at the Fermi level are occupied in the free
  system. Then, the gap observed at half-filling cannot be a spurious
  `shell' effect, even on small size systems.

A collection of chemical potential sweeps such as that in
Fig.~\ref{fig:gaps}, for different values of $\omega_4$, yields the
boundaries of the CDW region in a phase diagram in the ($\mu$,
$\omega_4$) plane.  We delimit the CDW region with the value of $\mu$
for which $1-\delta < \rho < 1+\delta$, using a small threshold value
$\delta$.  Figure \ref{fig:phasediagram_w05} shows $\omega=0.5$ in
panel (a) and $\omega=1$ in panel (b), for $\delta=0.05$.  The effect
of $\omega_4$ on the width and the position of the CDW gap is much
stronger at $\omega=0.5$ than at $\omega=1$, as the $\omega$
dependence in expression for the relative size of the anharmonic term
$\eta$ in Eq.~\ref{eq:eta} would suggest should be the case.  In both
cases, we observe a shift of the chemical potential at half-filling
towards smaller absolute values.  This shift can be explained
qualitatively using a simple approximation presented in Appendix
\ref{app:chem}. The red triangles in Fig. \ref{fig:phasediagram_w05}
show the values of $\mu$ at $\rho=1$ obtained with this approximation.
In both cases, $\omega=1$ and $\omega=0.5$, we observe a
  reduction of the charge gap (width of the CDW lobe) as $\omega_4$ is
  increased although the effect is more dramatic for $\omega=0.5$ (see
  Fig. \ref{fig:phasediagram_w05} (c)). The sensitivity of the system
  to the anharmonic term in the $\omega=0.5$ case is noticeable with
  strong differences already obtained for $\omega_4$ of order
  $10^{-3}$, a value for which $\eta \sim 0.5$. As the charge gap is
  much reduced for $\omega=0.5$, it becomes smaller than the
  $\omega=1$ charge gap as $\omega_4 \ge 0.0075$, despite the fact
  that it is much larger at small $\omega_4$
  (Fig. \ref{fig:phasediagram_w05} (c)).

\begin{figure}
\centerline{\includegraphics[width=8.5cm]{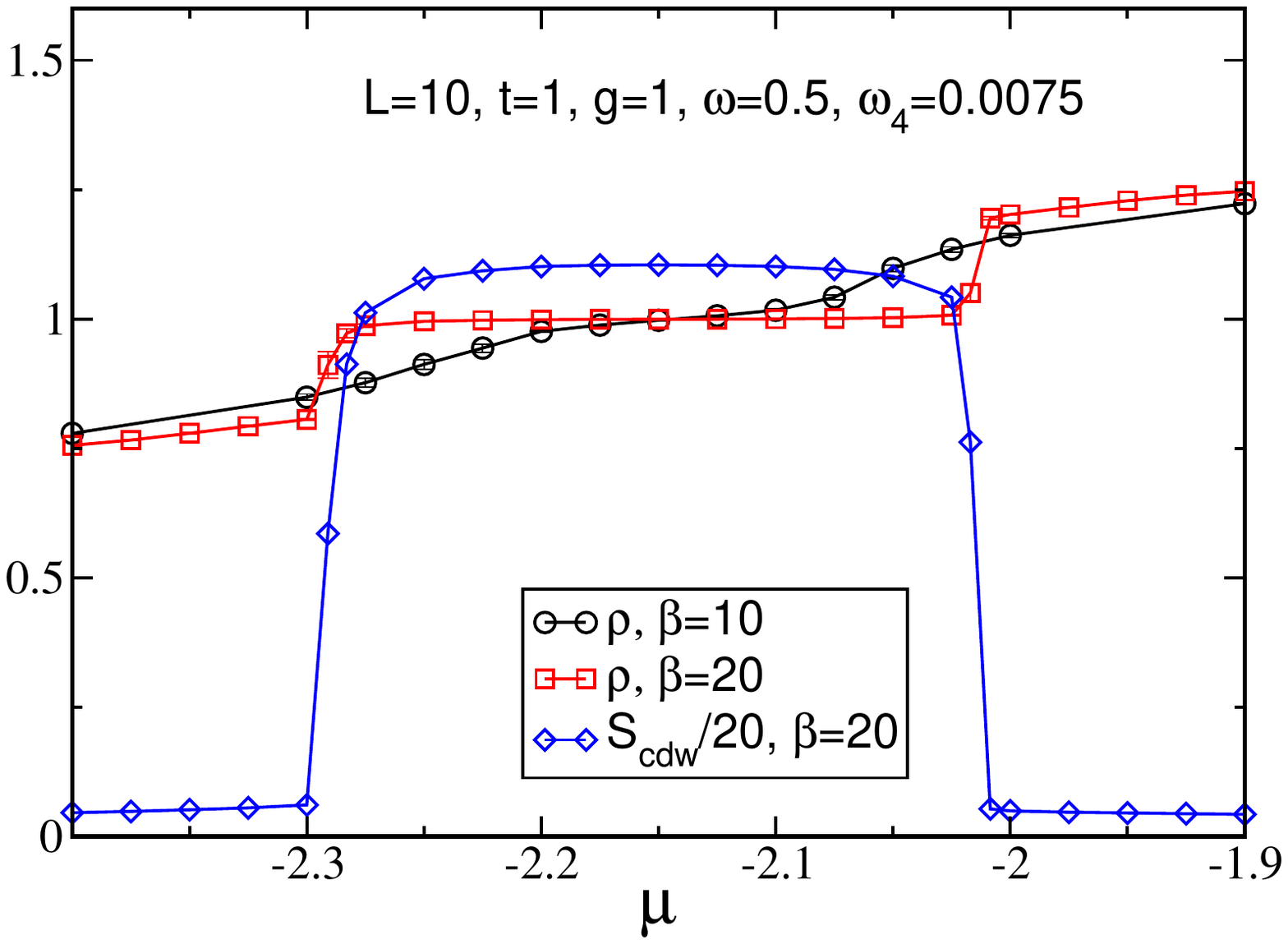}}
\caption{(Color online). The density, $\rho$, and CDW structure
  factor, $S_{\rm cdw}$, (rescaled for better visibility) for an
  $L=10$ system, focusing on the CDW plateau. Notice that, for
  $\beta=10$, the density does not yet show a plateau at half filling;
  $\beta=20$ is necessary for the system to display the ground state
  behavior and exhibit the CDW gap.  The structure factor $S_{\rm
    cdw}$ displays an abrupt change of values when the system is doped
  away from half-filling.
\label{fig:gapandS}}
\end{figure}

For large $\omega_4$, the gap becomes small in the $\omega=0.5$ case
(Fig. \ref{fig:phasediagram_w05} (c)). We verified for larger systems
that the small gap is not a finite size effect
(Fig.~\ref{fig:gapandS}). In these cases the plateau is rounded
(Fig.~\ref{fig:gapandS}) by thermal excitations at $\beta= 10$ and
inverse temperatures up to $\beta=20$ are needed to observe the flat
plateau typical of the ground state behavior. We also observe an
abrupt change of the density and of the CDW structure factor when the
system is doped away from half-filling. In the parameter
  regime which is accessible to our QMC, we did not find a value for
  $\omega_4$ where the gap at half-filling vanishes completely. The
  QMC simulations for $\omega_4 > 0.01$ become prohibitively difficult
  because they require exceedingly large values of $\beta$.

\begin{figure}
\centerline{\includegraphics[width=8.5cm]{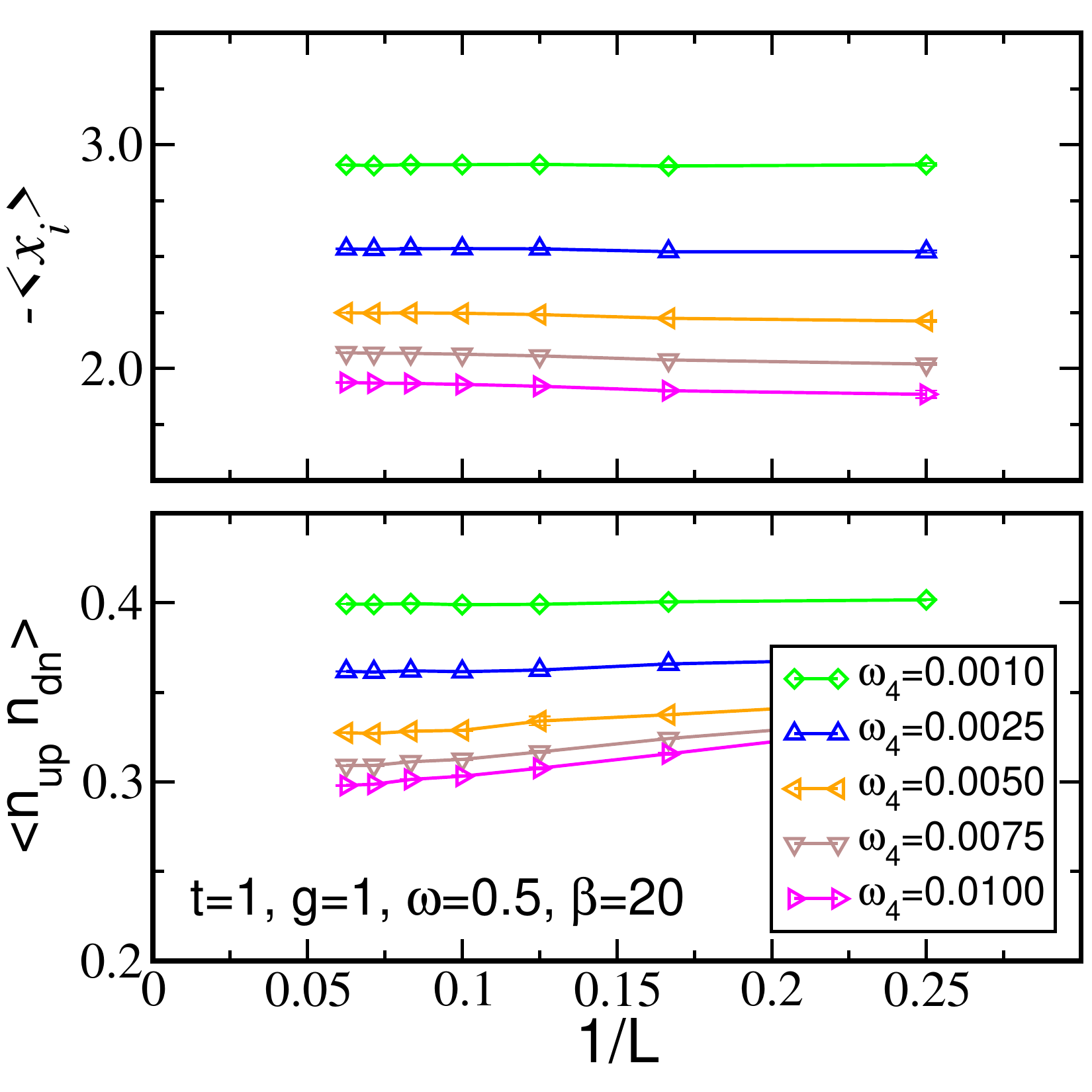}}
\caption{(Color online). Average value $\langle x_i\rangle $ (top) and
  double occupancy (bottom) as functions of $L^{-1}$ for different
  $\omega_4$ in half-filled systems. $|\langle x_i\rangle|$ is reduced
  as $\omega_4$ increases although it always extrapolates to nonzero
  values. For the harmonic case $\omega_4=0$, $\langle x_i \rangle = -4$. 
   The double occupancy is also reduced but always
  extrapolates to values larger than $\langle
  n_{i\uparrow}\rangle\langle
  n_{i\downarrow}\rangle=0.25$.\label{fig:xhf}}
\end{figure}

Knowing the values of $\mu$ where the system is half-filled, we
performed several targeted simulations at half-filling. In
Fig.~\ref{fig:xhf} (top), we show the evolution of the average value
$|\langle x_i\rangle|=-\langle x_i\rangle$ for different $\omega_4$
and sizes, $L$, in the CDW phase at half-filling for the $\omega=0.5$
case. We have not been able to obtain reliable results at this low
temperature and large sizes for small values of $\omega_4$, the
$\omega_4=0$ case being particularly difficult.  We then compare to
the analytical value at $\omega_4=0$, $|\langle x_i\rangle| =
|-\lambda/\omega^2| = 4$ (see App. \ref{app:chem}).  We find that,
although it always extrapolates to a nonzero value, $|\langle
x_i\rangle|$ is strongly reduced as $\omega_4$ increases, by a factor
of 2 at $\omega_4=0.01$ compared to $\omega_4=0$.  This is expected as
the anharmonicity penalizes large values of $x$, as does a large value
of $\omega$. As the phonon field generates an effective attraction
between the fermions, this attraction is weakened and the double
occupancy $\langle n_{i\uparrow}n_{i\downarrow}\rangle$ is
correspondingly reduced, although it always remains larger than the
uncorrelated value $\langle n_{i\uparrow}\rangle \langle
n_{i\downarrow}\rangle$.  This suppression of $|\langle x_i\rangle|$
and the resulting reduction of the effective attraction between
fermions explain the observed shrinking of the CDW charge gap.

For $\omega=1$, we find the same effects but with a much reduced
amplitude. In that case, for $\omega_4=0$, $|\langle x_i\rangle|=\sqrt{2}$ which we have 
confirmed numerically (for $L=16$ and $\beta=20$, we find $|\langle x_i\rangle|
= 1.4143(4)$). $|\langle x_i\rangle|$ varies from $\sqrt{2}$ down to $|\langle
x_i\rangle|\simeq1.3$ when $\omega_4$ varies from $0$ to $0.005$ while
$\langle n_{i\uparrow}n_{i\downarrow}\rangle$ decreases from $\langle
n_{i\uparrow}n_{i\downarrow}\rangle\simeq 0.35$ to $\langle
n_{i\uparrow}n_{i\downarrow}\rangle\simeq 0.33$ over the same
$\omega_4$ interval.

\begin{figure}
\centerline{\includegraphics[width=8.5cm]{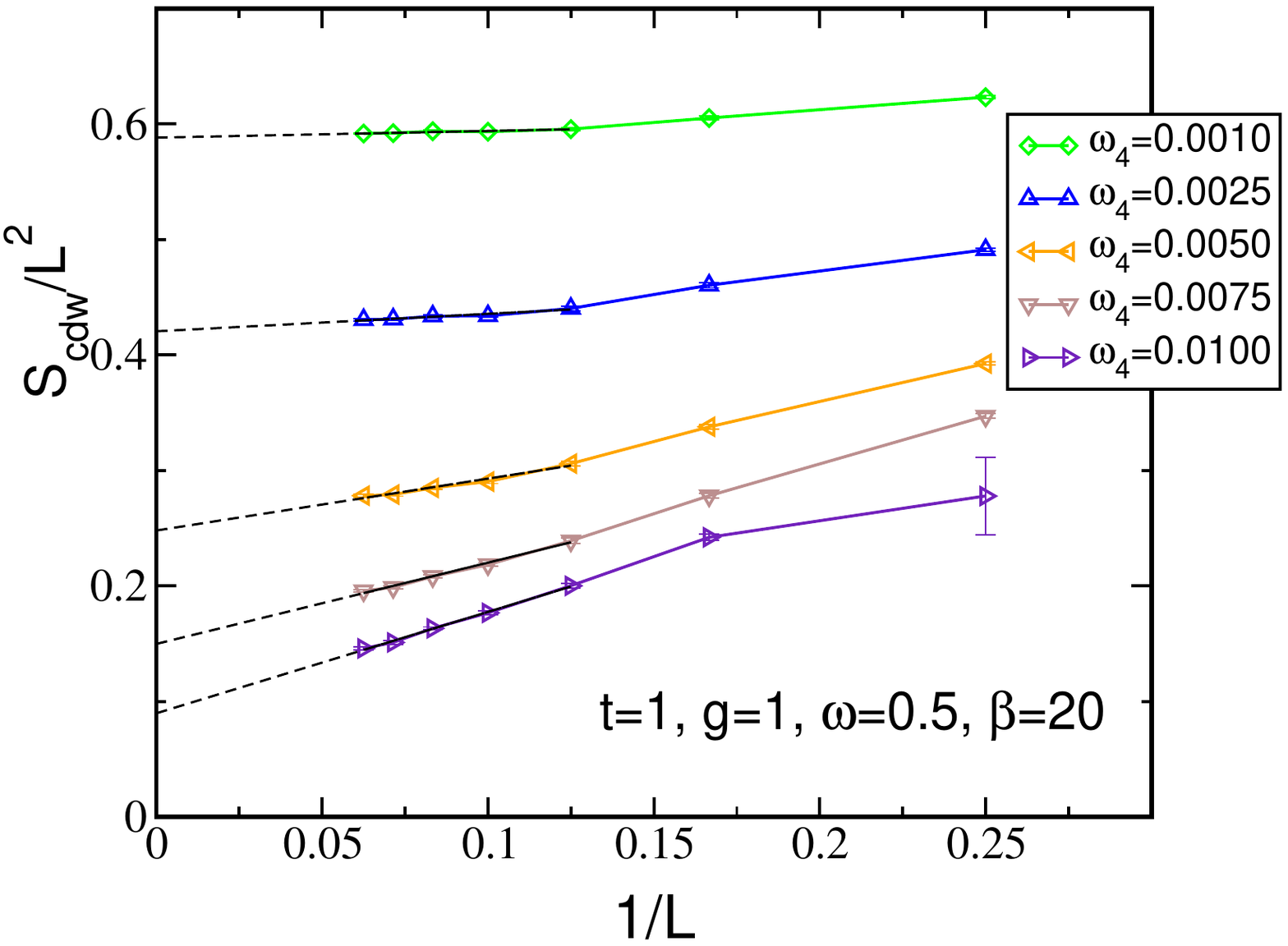}}
\caption{(Color online). Structure factor $S_{\rm cdw}$ as a function
  of size $L$ for different values of $\omega_4$ at half-filling. For
  all $\omega_4$, the linear extrapolation of $S_{\rm cdw}$ to $L
  \rightarrow \infty$ is nonzero. The linear extrapolation is based on
  fits for the data with $L\ge 8$.
\label{fig:Scdw}}
\end{figure}

The gapped phase is expected to show CDW order, which we confirmed by
a direct study of $S_{\rm cdw}$ at half-filling for different
$\omega_4$ (Fig.~\ref{fig:Scdw}).  In all the cases studied here,
$S_{\rm cdw}$ extrapolates to a nonzero value in the thermodynamic
limit $L \rightarrow \infty$ and is reduced as $\omega_4$ increases.
We verified the presence of a corresponding CDW order in
  the distribution of $|\langle x_i\rangle|$.

\subsection{Finite temperature transition to CDW order}

To complete this analysis of the CDW behavior at half-filling, we
analyze the transition to this phase as the temperature, $T$, is
lowered. The CDW transition breaks translation symmetry between the
two sublattices of the square lattice. It is, therefore, in the
universality class of the two-dimensional Ising model with a finite
critical temperature, $T_c$, and 2D Ising critical exponents.

\begin{figure}
\centerline{\includegraphics[width=8.5cm]{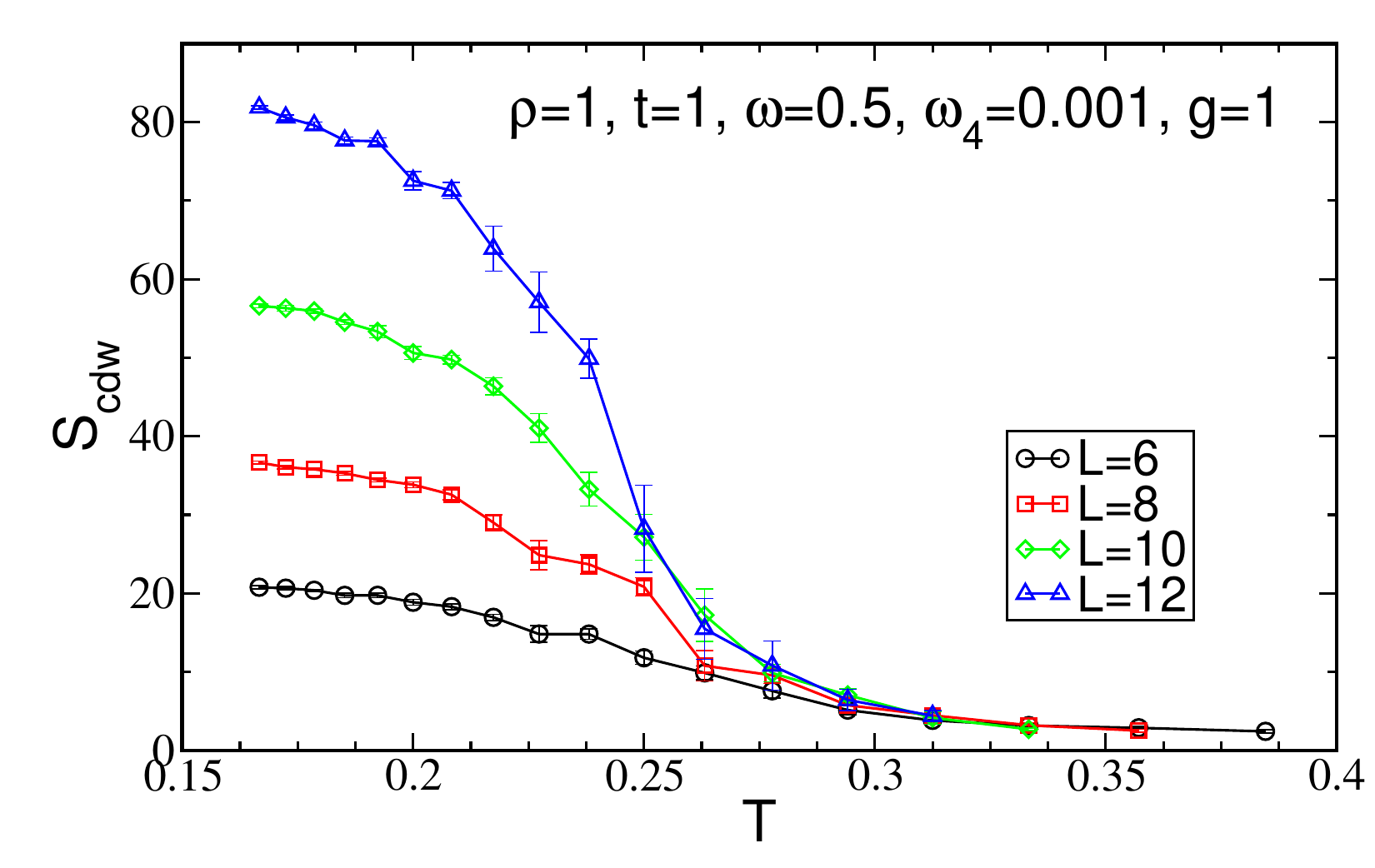}}
\centerline{\includegraphics[width=8.5cm]{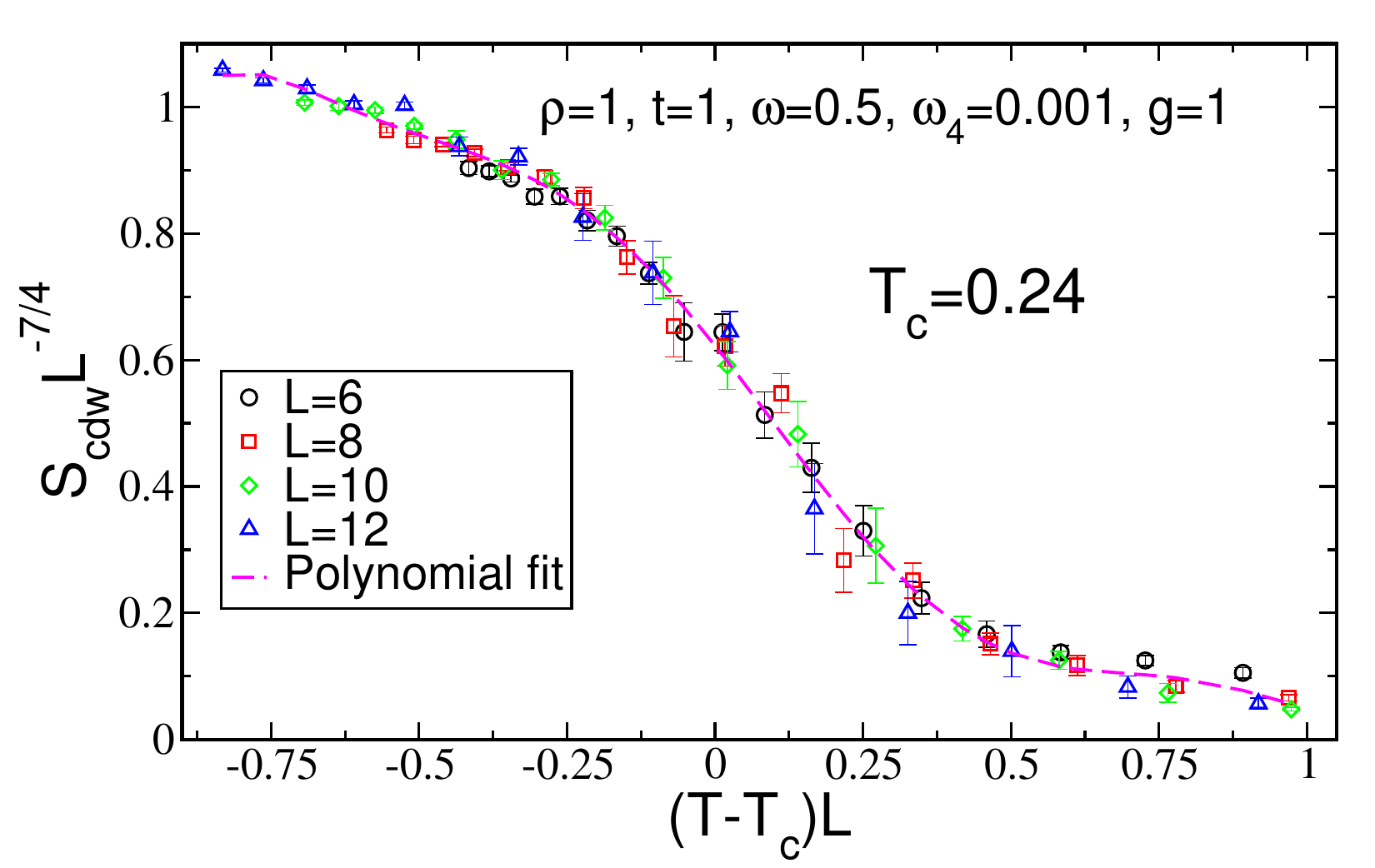}}
\caption{(Color online). Finite size analysis for $\omega=0.5$ and
  $\omega_4=0.001$. (Top) Structure factor for different sizes as a
  function of temperature $T$. (Bottom) Rescaled structure factor as a
  function of the rescaled reduced temperature. The critical
  temperature $T_c = 0.24 \pm 0.01$ is chosen to obtain the best possible collapse
  between the different curves.}
\label{fig:scaling}
\end{figure}

We used standard finite size scaling analysis where, close to the
transition, the structure factor behaves as
\begin{equation}
\frac{S_{\rm cdw}}{L^2} = L^{-2\beta/\nu} \tilde
S(L^{1/\nu}t) \label{eq:scaling} \Rightarrow S_{\rm cdw} = L^{7/4}
\tilde S(Lt),
\end{equation}
with the critical exponents $\beta=1/8$ and $\nu=1$, $t=T-T_c$ is the
reduced temperature, and $\tilde S$ is a universal scaling
function. As the critical exponents are known a priori, the only
unknown quantity is $T_c$ which will be chosen to optimize the
superposition of the curves obtained for different system sizes (see
Fig. \ref{fig:scaling}).  To do so, we choose a value of $T_c$,
rescale all the data according to Eq. \ref{eq:scaling} and fit those
data with a high degree polynomial. We then determine the optimal
value of $T_c$ as the one that minimizes the distance between the
polynomial fit and the data.  At large $\omega_4$, finite size
corrections to scaling are larger and we have not found sizes where
finite size scaling analysis can be used (see Appendix~\ref{app:ffs}),
which limits the range in which we are able to determine the critical
temperature.

\begin{figure}
\centerline{\includegraphics[width=8.5cm]{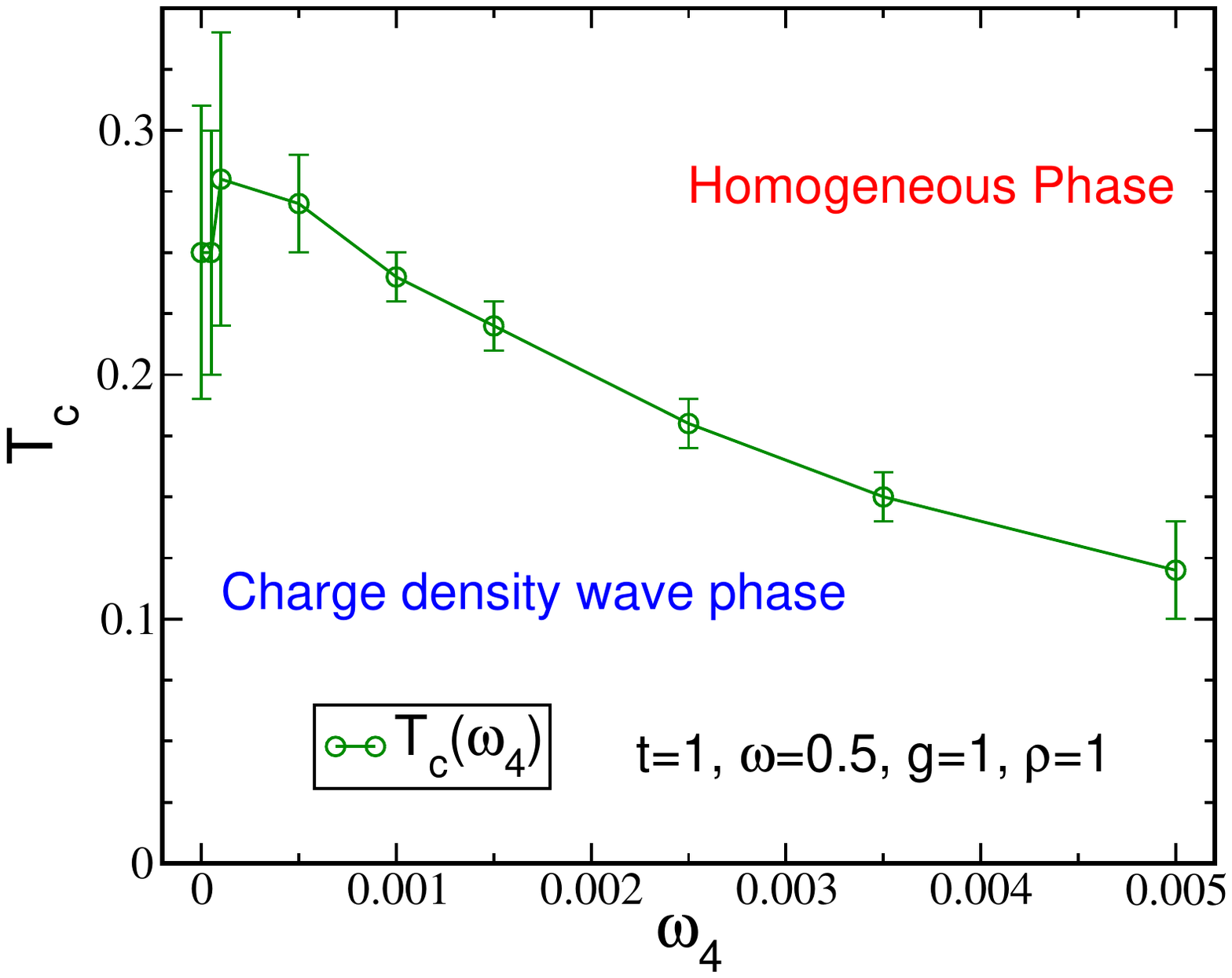}}
\caption{(Color online). CDW critical temperature, $T_c$, at
  half-filling as a function of anharmonicity, $\omega_4$, for
  $\omega=0.5$, $t=1$, $g=1$.  With sizes up to $L=12$, the finite
  size analysis was only feasible for $\omega_4 \le 0.005$.
\label{fig:Tcw1}}
\end{figure}

As expected, $T_c$ decreases with $\omega_4$ (see
Fig. \ref{fig:Tcw1}). This behavior could have been inferred from the
evolution of the charge gap and the structure factor at low
temperatures.  Compared to the infinite dimension results presented in
Ref.[\onlinecite{Freericks96}], which focused on the $\omega=0.5$
case, we observe a similar reduction of $T_c$ with $\omega_4$.  Our
simulations show that the critical temperature changes from $T_c
\simeq 0.25$ at $\omega_4=0$ down to $T_c \simeq 0.12$ at $\omega_4 =
0.005$.  Freericks {\it et al.}\cite{Freericks96} also predicted an
initial increase of $T_c$ with $\omega_4$.
While we observe such an
effect in some simulations, we cannot give a definite conclusion
concerning this increase of $T_c$ due to the lack of precision of our
data for small $\omega_4$. The most remarkable difference with the
infinite dimension description is the range of $\omega_4$ over which
noticeable changes are observed: we found a strong modification of
critical temperature for $\omega_4 \simeq 5\cdot10^{-3}$ whereas
similar variations are found in \cite{Freericks96} for $\omega_4
\simeq 10^{-1}$.

In the $\omega=1$ case, for a similar range of $\omega_4$, we did not
observe a strong change of the value of the critical temperature with
$\omega_4$.  Values of $\omega_4$ where we could apply the finite size
analysis were more restricted than for $\omega=0.5$ and we could only
get results for $\omega_4$ up to 0.001.  At $\omega_4=0$, we found a
critical temperature of $T_c =0.16 \pm 0.01$ which is compatible with
values found recently in similar cases \cite{Costa2017, Costa2018,
  Weber2018}. For $\omega_4=0.001$ the critical temperature is barely
reduced to $T_c = 0.15 \pm 0.01$. This was expected as we observed
that, in this case, the anharmonicity hardly changes the width of the
gap at half-filling.

Finally, for both $\omega=0.5$ and $\omega=1$, we
  observed a reduction of the charge gap and critical temperature as
  $\omega_4$ increases but we did not observe a disappearance of the
  CDW phase in the accessible parameter range. For larger values of
  $\omega_4$, as in the pure Holstein case \cite{Hohenadler19}, there
  are two possible scenarios. The first is a persistence of the CDW
  phase at half-filling with decreasing gap and critical temperature,
  which is possible because our model retains the Fermi surface
  nesting present in the Holstein model that favors CDW order.  The
  second scenario is the existence of a critical value of $\omega_4$
  above which the CDW phase no longer exists.

\section{Doped system}

\subsection{First order transition near half-filling}

\begin{figure}
\centerline{\includegraphics[width=8.5cm]{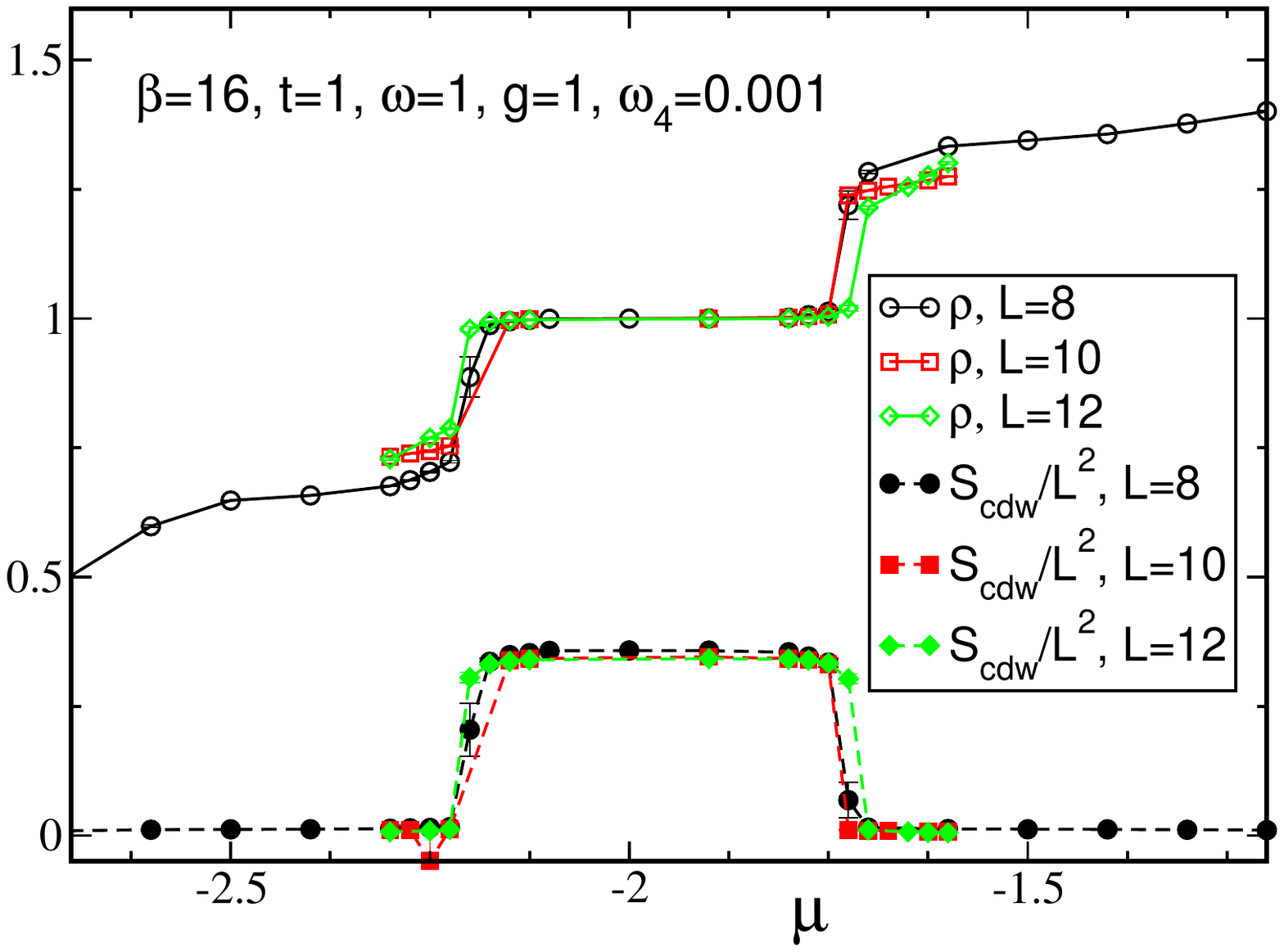}}
\caption{(Color online). Density, $\rho$, and structure factor,
  $S_{\rm cdw}$, as functions of the chemical potential for $\omega=1$
  and several sizes at $\beta=16$.  An abrupt change of the density is
  found when the system is doped away from half-filling, which also
  corresponds to the disappearance of CDW order.
\label{fig:jump}}
\end{figure}

\begin{figure}
\centerline{\includegraphics[width=8.5cm]{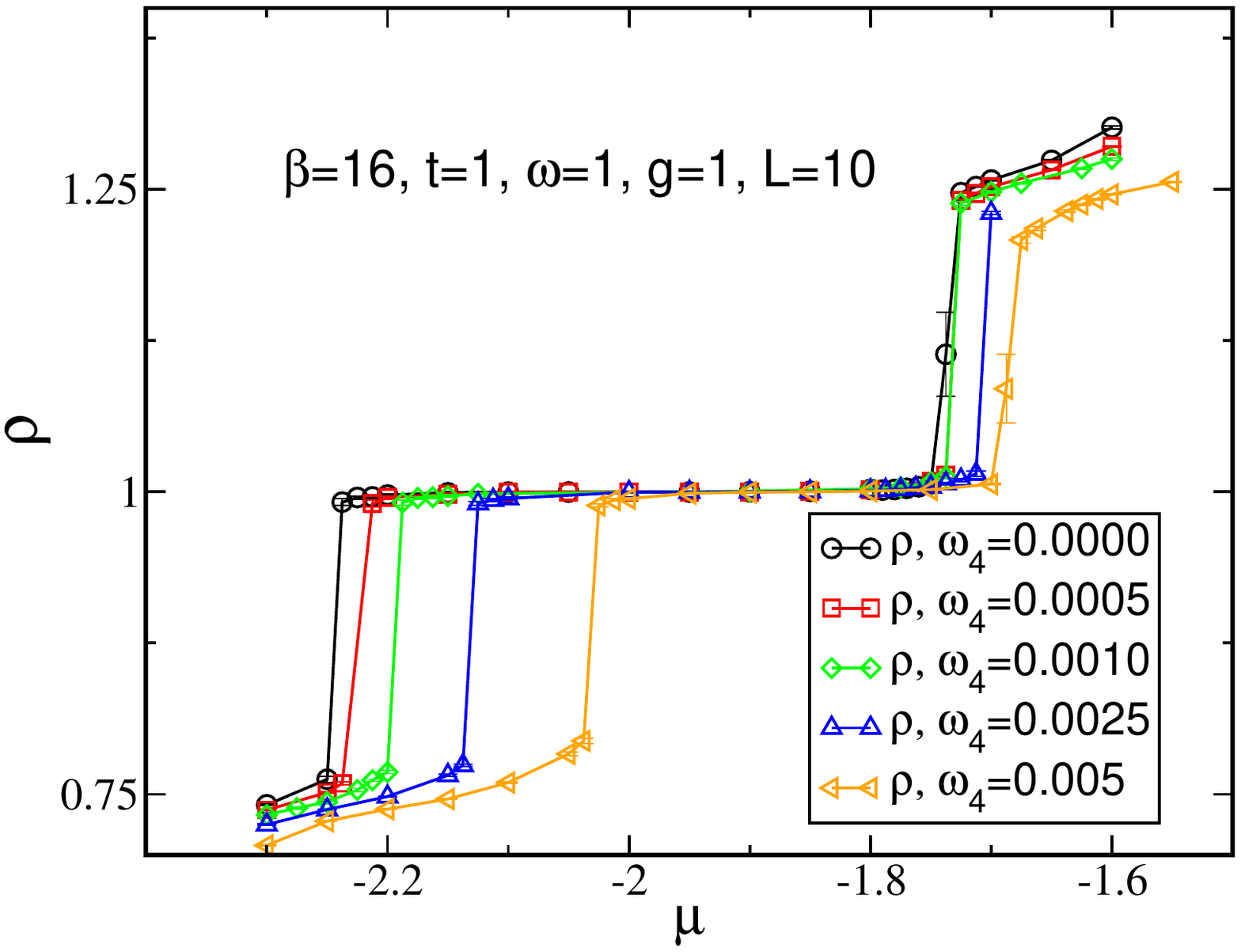}}
\caption{(Color online). Density, $\rho$, as a function of the
  chemical potential for $\omega=1$ and several values of $\omega_4$
  at $\beta=16$.  The abrupt change of the density found when the
  system is doped is present for all values of $\omega_4$.
\label{fig:jump2}}
\end{figure}

The infinite dimension prediction by Freericks {\it et. al.}
\cite{Freericks96} shows a CDW when the system is doped away from
half-filling as well as a SC phase. At sufficiently low temperature,
in our Langevin simulations, the evolution of the density with $\mu$,
for both $\omega=0.5$ (Fig.~\ref{fig:gapandS}) and $\omega=1$
(Fig.~\ref{fig:jump}), exhibits an abrupt change of the density in the
neighborhood of the CDW plateau. We see that these jumps are not
finite size effects as their amplitude does not vary much with the
size of the system (Fig.~\ref{fig:jump}). We observe this kind of
discontinuity in the density for all values of $\omega_4$, down to
$\omega_4=0$.  They are more pronounced for the lower phonon frequency
$\omega=0.5$. Below half-filling, for $\omega=1$, the density jumps
from $\rho\simeq 0.75$ to $\rho=1$ and the extent of the jump does not
depend much on $\omega_4$ (Fig.~\ref{fig:jump2}) although it decreases
slightly with increasing $\omega_4$. We observe a similar jump above
half-filling.

For $\omega=0.5$, the finite temperature effects are stronger and it
is more difficult to assess precisely the size of the discontinuity.
It appears the change is from $\rho=1$ to a value which is around
$\rho\simeq 0.25$ for $\omega_4=0.001$ whereas, as can be observed in
Fig.~\ref{fig:gapandS}, the jump is reduced to $\rho=1$ down to $\rho
\simeq 0.75$ for $\omega_4=0.0075$.  The structure factor is
essentially zero when the density is no longer one. We do not find, at
these low temperatures, any sign of an intermediate doped region with
nonzero structure factor.

\begin{figure}
\centerline{\includegraphics[width=8.5cm]{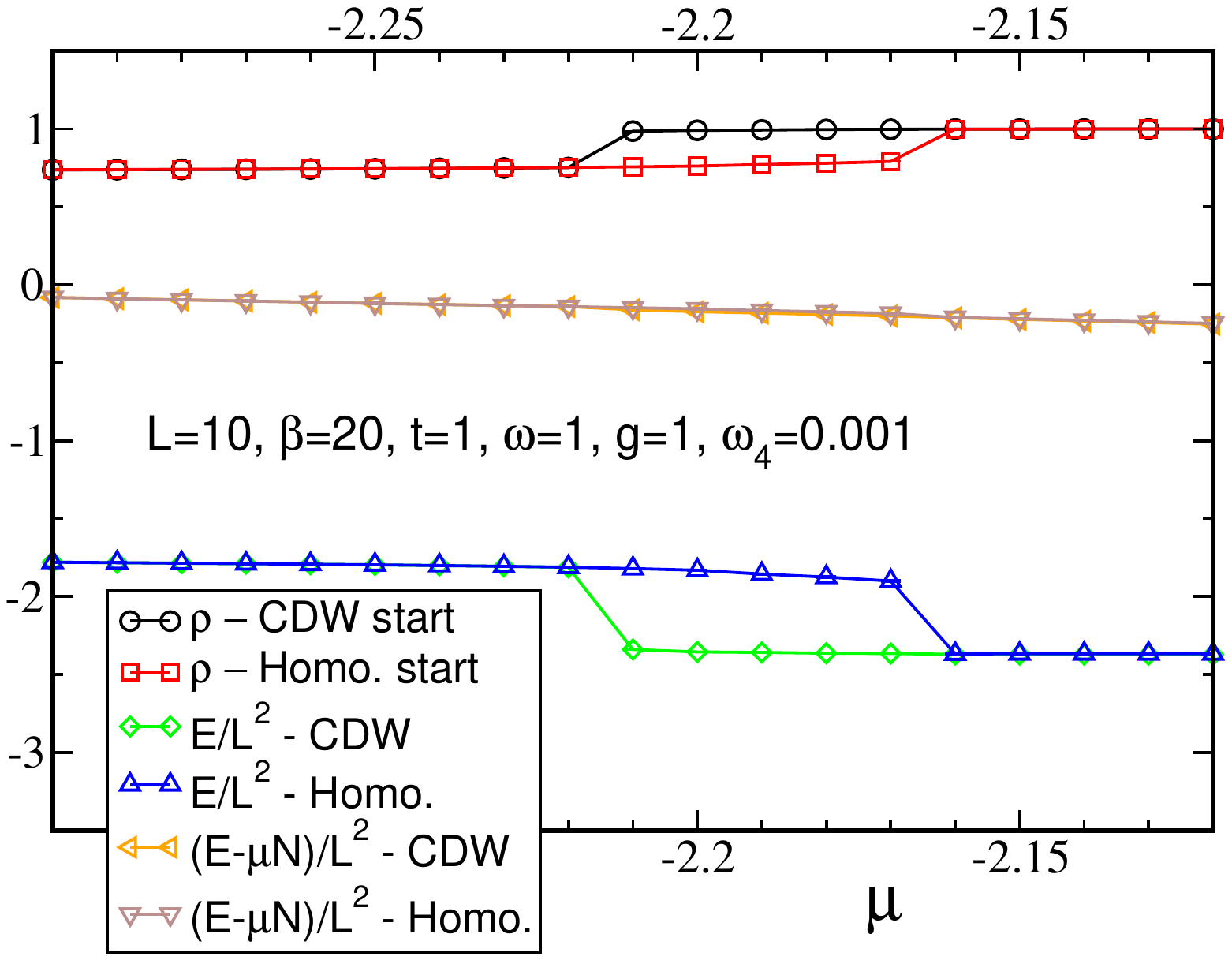}}
\caption{(Color online). Density, $\rho$, energy per site, $E/L^2$,
  and grand potential per site, $E/L^2-\mu \rho$, as functions of
  $\mu$ for different initial conditions of the simulations (CDW start
  or homogeneous start). We observe hysteresis with an intermediate
  region where two different phases coexist.  \label{fig:hysteresis}}
\end{figure}

Such discontinuities indicate that the transition, as $\mu$ is
changed, is of first order. If the simulations were done in the
canonical ensemble, there would be phase separation between a CDW and
a uniform phase in the jump region, as was observed in bosonic Hubbard
models when the system is doped away from a CDW phase
\cite{Batrouni2000}. Such a transition was recently observed in
variational Monte Carlo simulations \cite{Ohgoe17} and was also
reported in [\onlinecite{bradley20}].  To confirm the first order
nature of the transition, we analyzed the behavior of the density and
energy for a large enough system, $L=10$, at low temperature,
$\beta=20$, doping below half-filling (Fig. \ref{fig:hysteresis}).  By
choosing appropriate values of the phonon coordinates, we are able to
start the Langevin simulations with two different initial conditions:
a homogeneous solution and a CDW one.

For such large systems, the simulations remain ``stuck" in the kind of
phase that was initially imposed upon the system indicating a
metastability characteristic of first order transitions. This leads to
hysteresis as is seen clearly in Fig.~\ref{fig:hysteresis}.  In the
hysteresis region, we find that the grand potentials $E-\mu N$ of the
two phases to be essentially equal, and since $E-\mu N$ is minimized
at equilibrium, we then observe two equivalent solutions in this
chemical potential range. As a consequence, the energy $E$ of the CDW
phase is lower than that of the homogeneous phase in the coexistence
region.

Contrary to what was observed in infinite dimension
\cite{Freericks96}, we do not find in two dimensions a region away
from half-filling where CDW order survives.  It is noticeable that
smaller systems, such as the ones used at the beginning of this study
(Fig.~\ref{fig:gaps}), or higher temperatures may give the false
signal that there is CDW away from half filling because it is possible
to choose a value of $\mu$ that gives an average density located in
the unstable region. The system will then have a broad density
distribution ranging from the low homogeneous phase density up to
$\rho=1$, and since measured quantities are averaged over this wide
distribution, the structure factor can appear to be nonzero
\cite{Batrouni2000}.

\subsection{Superconducting behavior}

Away from half-filling, the system is expected to become
superconducting at low temperatures. However, in general, the
transition temperatures appear to be low \cite{esterlis18, bradley20}.
For the $\omega=1$ harmonic Holstein model, the transition towards a
SC state happens for $\beta \simeq 28$ \cite{bradley20} and even
larger inverse temperature for $\omega=0.5$. This makes it difficult
to observe the effects of anharmonicity on the critical temperature
itself, especially since $\omega_4$ is expected to reduce $T_c$ even
further.  Instead, we will focus on the evolution of the
superconducting susceptibility $\chi_s$, without attempting to discern
where it might diverge.

\begin{figure}
\centerline{\includegraphics[width=8.5cm]{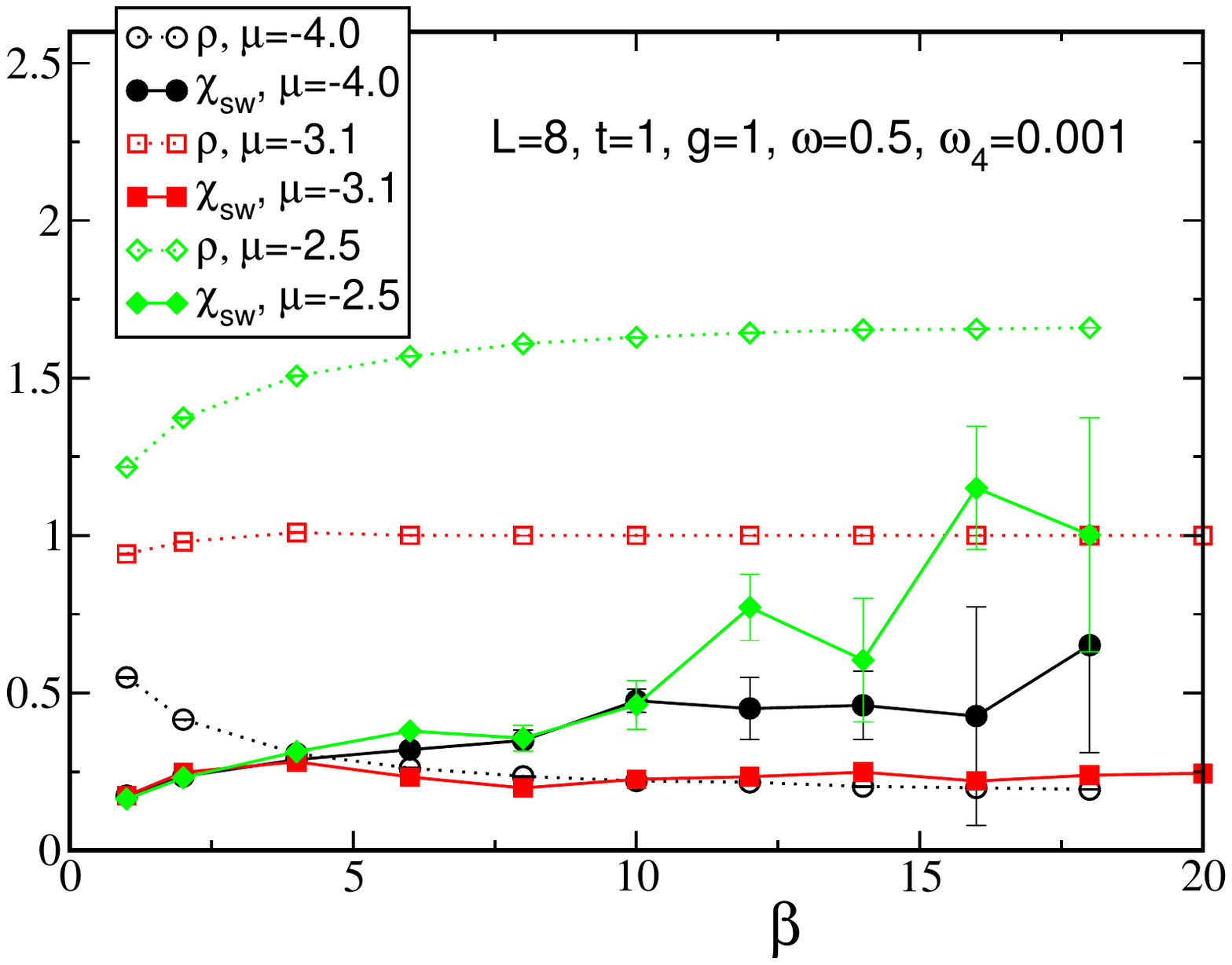}}
\caption{(Color online). Density, $\rho$, and s-wave pairing
  susceptibility, $\chi_{\rm s}$, as functions of $\beta$ for
  $\omega=0.5$, $\omega_4=0.001$, and $g=t=1$. $\mu=-3.1$ corresponds
  to the half-filled system, $\mu=-4$ to a lower density $\rho \simeq
  0.25$, and $\mu=-2.5$ to $\rho>1.5$.
\label{fig:dens_susc}}
\end{figure}

Fig.~\ref{fig:dens_susc} shows the evolution of the density, $\rho$,
as well as the superconducting susceptibility, $\chi_{\rm s}$, as
functions of $\beta$ for $\omega=0.5$ and three values of $\mu$
corresponding to densities below, at, and above half-filling. We first
observe that, away from half-filling, the density reaches its ground
state behavior only above $\beta=10$.  As can be expected, the pairing
susceptibility at half filling does not diverge, but remains small.
For the doped system, it was not possible to observe a divergence of
$\chi_{\rm s}$ in the range of accessible temperatures. As shown in
Fig.~\ref{fig:dens_susc}, statistical fluctuations in $\chi_{\rm s}$
become large in the doped system for $\beta > 10$, and would become
even more problematic in attempting to approach the superconducting
transition one expects at much lower temperature.

\begin{figure}
\centerline{\includegraphics[width=8.5cm]{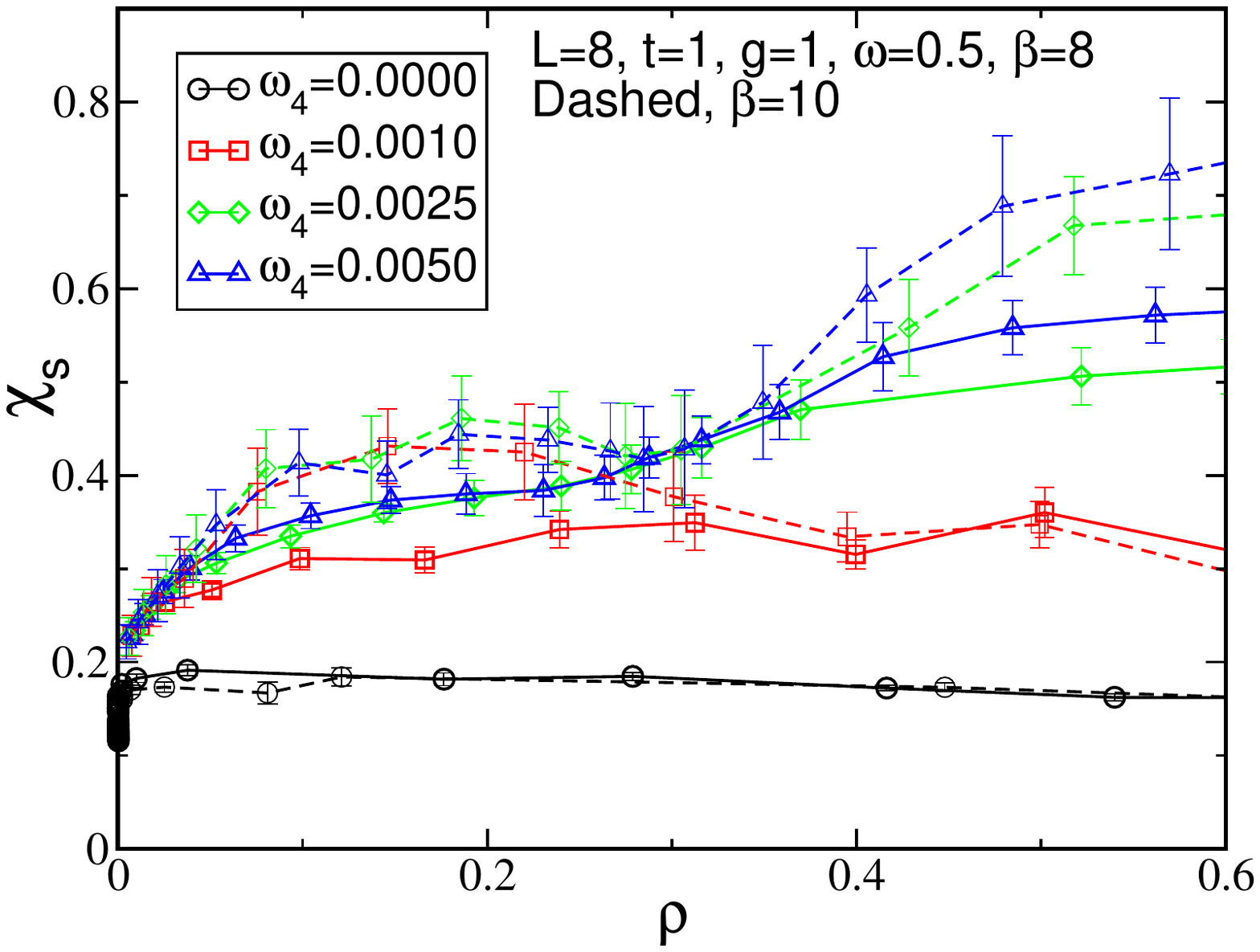}}
\caption{(Color online). The s-wave pairing susceptibility, $\chi_{\rm
    s}$, as a function of density, $\rho$, for $\omega=0.5$, $g=t=1$, different values of $\omega_4$, and $\beta=8,10$. The susceptibility
  increases as $\omega_4$ or $\beta$ increases.
\label{fig:SCw05}}
\end{figure}

With this limited access to superconducting behavior, we study the
effects of the anharmonicity through the evolution of $\chi_{\rm s}$
as a function of $\omega_4$ for small size and intermediate
temperatures $\beta=8,10$.  In Fig.~\ref{fig:SCw05}, $\omega=0.5$, we
observe that the superconducting response increases rapidly as
$\omega_4$ is increased for a density range $0<\rho\leq 0.6$. Once
again, we observe that increasing $\omega_4$ has roughly the same
effect as increasing $\omega$; it promotes superconductivity. We did
not study the region between $\rho=0.6$ and $\rho=1$ as it corresponds
to the unstable region between homogeneous and CDW phases. We remark
that, for the smaller values of $\omega_4$, the system will already be
unstable for $\rho > 0.25$.

\begin{figure}
\centerline{\includegraphics[width=8.5cm]{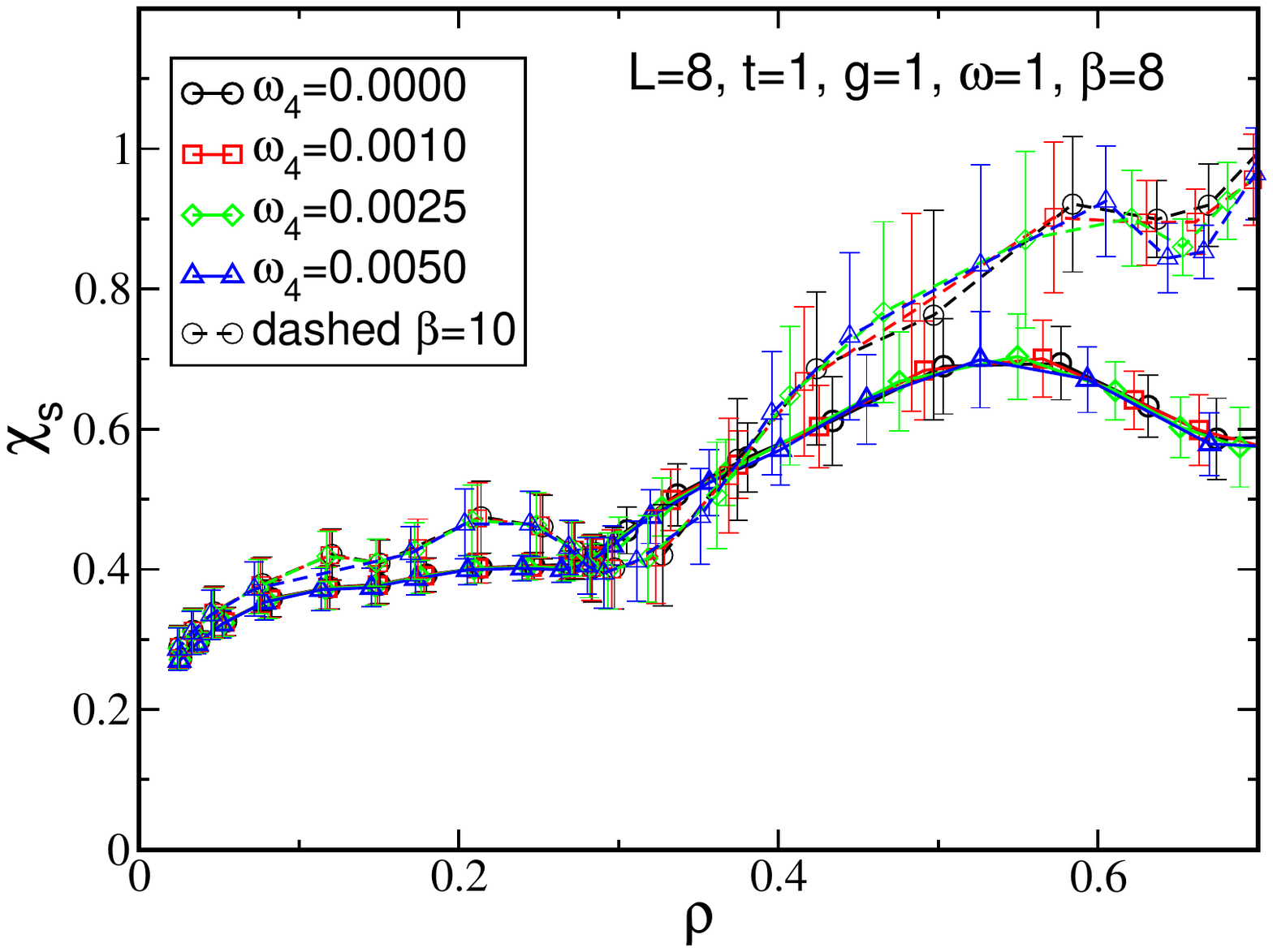}}
\caption{(Color online). The s-wave pairing susceptibility,
$\chi_{\rm s}$, as a function of density, $\rho$,
  for $\omega=1$, $g=t=1$, different values of $\omega_4$, and $\beta=8,10$. The
  susceptibility is not sensitive to changes in $\omega_4$ but
  increases with $\beta$. \label{fig:SCw1}}
\end{figure}

For $\omega=1$, the anharmonicity has limited effect on the SC
susceptibility at the values of $\omega_4$ we studied
(Fig. \ref{fig:SCw1}).  This parallels the small $\omega_4$ dependence
of the CDW lobe in the phase diagram
(Fig.~\ref{fig:phasediagram_w05}(b)) for this phonon frequency.

In both cases, $\omega=0.5$ (Fig.~\ref{fig:SCw05}) and $\omega=1$
(Fig.~\ref{fig:SCw1}), we observe an increase of $\chi_{\rm s}$ as
$\beta$ goes from $8$ to $10$ but are not able to observe the
divergence of $\chi_{\rm s}$.

\section{Conclusions}

In this work, we studied the effect of an anharmonic quartic term on
the physics of the Holstein model at strong electron-phonon coupling,
$g=1$, and phonon frequencies $\omega=0.5$ and $\omega=1$.  We
observed similar effects of the anharmonicity for the two phonon
frequencies, but the effects were much reduced for the $\omega=1$ case
in the range of anharmonicities we studied.  We found that the main
effect of the quartic term is to reduce the importance of the
electron-phonon coupling compared to the phonon potential energy. At
half-filling, this shrinks the charge gap and leads to a suppression
of the CDW structure factor at zero temperature and to a lowering of
the critical temperature for the CDW transition.

The behavior of the density as one approaches an insulating plateau
has been a central interest in a number of contexts, including early
Bethe {\it ansatz} solutions of the 1D fermion Hubbard
model\cite{frahm90}.  For the 2D fermion Hubbard model, Assaad and
Imada used QMC methods to extract critical exponents\cite{assaad96}.
Further fermion work is reviewed in [\onlinecite{imada98}].  In
parallel, similar issues have been central to the investigation of the
boson-Hubbard model, including theoretical prediction\cite{fisher89}
of the mean field nature of the density controlled transition into the
Mott lobe which were confirmed by QMC\cite{batrouni90}.

In this work, we have added further information to this area by
studying the anharmonic Holstein Hamiltonian.  Doping the system away
from half-filling, we observed a first order phase transition between
the CDW phase at half filling and a homogeneous phase at lower
densities. This first order transition is present, though not widely
studied previously, in the harmonic Holstein model
\cite{Ohgoe17,bradley20}.

In the homogeneous phase below half-filling, for $\omega=0.5$, we
observed a clear enhancement of the superconducting susceptibility at
finite temperature as $\omega_4$ is increased. However, with the
limited range of accessible temperatures, we were not able to observe
the superconducting transition.  For $\omega=1$, $\omega_4$ does not
have a strong effect on the superconducting response.

The results we presented here show that the transitions from $\rho=1$
to $\rho>1$ and to $\rho<1$, as $\mu$ is tuned, are both first order
(Fig.~\ref{fig:jump2}). However, most of our results for the doped
system focused on $\rho < 1$, and, since the system is no longer
particle-hole symmetric, it would be interesting to study its
properties above half filling further.  To complete the understanding
of the role of the quartic term, it is necessary to study the system
at other coupling parameters, in particular lower values of $g$.

\bigskip

\begin{acknowledgments}
We thank Owen Bradley for insights on the superconducting behavior of
these systems, and we thank Steve Johnston and Seher Karakuzu for very useful
discussions and comments. This work was supported by the French government, through
the UCAJEDI Investments in the Future project managed by the National
Research Agency (ANR) with the reference number ANR-15-IDEX-01 and by
Beijing Computational Science Research Center.  KB acknowledges
support from the center of Materials Theory as a part of the
Computational Materials Science (CMS) program, funded by the
U.S. Department of Energy, Office of Science.  The work of RTS was
supported by the grant DE‐SC0014671 funded by the U.S. Department of
Energy, Office of Science.  B.C-S acknowledges support from the
UC-National Laboratory In-Residence Graduate Fellowship through the UC
National Laboratory Fees Research Program.
\end{acknowledgments}

\appendix

\section{Approximate values of $\langle x_i \rangle$, $\mu$ and $U_{\rm eff}$ \label{app:chem}}

In the harmonic case, the value of the chemical potential at half-filling and of the average phonon displacement can 
be found exactly by a particle-hole transformation combined with a transformation of the phonon displacement
\begin{eqnarray}
&&c_{i\sigma} = (-1)^i \tilde{c}^\dagger_{i\sigma}, \quad
  c^\dagger_{i\sigma} = (-1)^i \tilde{c}_{i\sigma}, \nonumber \\ &&x_i
  = -\tilde{x}_i + x_0, \quad p_i = - \tilde{p}_i. \label{eq:transfo}
\end{eqnarray}
The transformed Hamiltonian is the same as the
original one provided that $x_0 = -2\lambda/\omega^2 = -2 \sqrt{2\omega} g / \omega^2$, which cancels
out terms that are linear in $x_i$, and that $\mu = \lambda x_0/2$, which is then the chemical potential
at half-filling.

Using this value of $\mu$, the resulting Hamiltonian can be expressed in terms
of $\delta_i = x_i - x_0/2$ and is invariant under the particle-hole transformation combined with a
$\delta_i \rightarrow -\delta_i$ transformation. This shows
that $\langle x_i\rangle$ is exactly equal to $x_0/2=-\lambda/\omega^2 = - \sqrt{2\omega} g / \omega^2$ in the ground state at half-filling.

We can roughly estimate the relative sizes of the harmonic and
anharmonic terms as follows: at half-filling, a CDW phase will
develop and we will, approximately, have an alternation of empty
and doubly occupied sites. As $\langle x_i\rangle = x_0/2$ when
averaged over all sites, the value of $x_i$ on doubly occupied sites
can be approximated by $x_0$.
If we then compute the ratio $\eta$ of
the anharmonic to harmonic terms at $x_0$ we obtain,
\begin{align}
\eta\equiv \frac{\omega_4 x_0^4}{\omega^2 x_0^2/2} = \frac{16 \omega_4
  \, g^2}{\omega^5}
\end{align}

We can also estimate the effective attraction between fermions. Completing the square of the phonon term
at $\omega_4=0$ results in
\begin{align}
\frac{1}{2} \omega^2 x^2 + \lambda x n =
\frac{1}{2} \omega^2 \left( x + \frac{\lambda n}{\omega^2} \right)^2 
-\frac{\lambda^2}{2 \omega^2} n^2
\label{eq:ueffandx0}
\end{align}
Since $n^2 = n_\uparrow+n_\downarrow + 2 n_\uparrow n_\downarrow $,
the second term on the right hand side of this expression gives an attractive
interaction between up and down electrons, $U_{\rm eff} =
-\lambda^2/\omega^2$.  The first term shows that the phonon potential
energy is indeed minimized at $x_0 = -2 \lambda/\omega^2$ on a doubly occupied site.

Adding the anharmonic term (Eq.$\,$\ref{eq:ham2}) breaks the
particle-hole symmetry and it is no longer possible to derive
analytically the value of the chemical potential at half-filling. One
can obtain an approximate value by using particle-hole transformation
Eq.~\ref{eq:transfo} and by canceling the terms that are linear in $x_i$, neglecting higher order
terms. This
leads to the following equation for $x_0$:
\begin{equation}
\omega^2 x_0  + 4\omega_4 x_0^3 = -2\lambda
\end{equation}
and the chemical potential at half-filling is approximately given by
$\mu = \lambda x_0/2$.  $|x_0|$ is obviously reduced as $\omega_4$
increases and, then, the chemical potential at half-filling is
increased. This approximate formula is used to derive the chemical
potential shown in Fig. \ref{fig:phasediagram_w05}.

\section{Corrections to finite size scaling \label{app:ffs}}

\begin{figure}[b]
\centerline{\includegraphics[width=8.5cm]{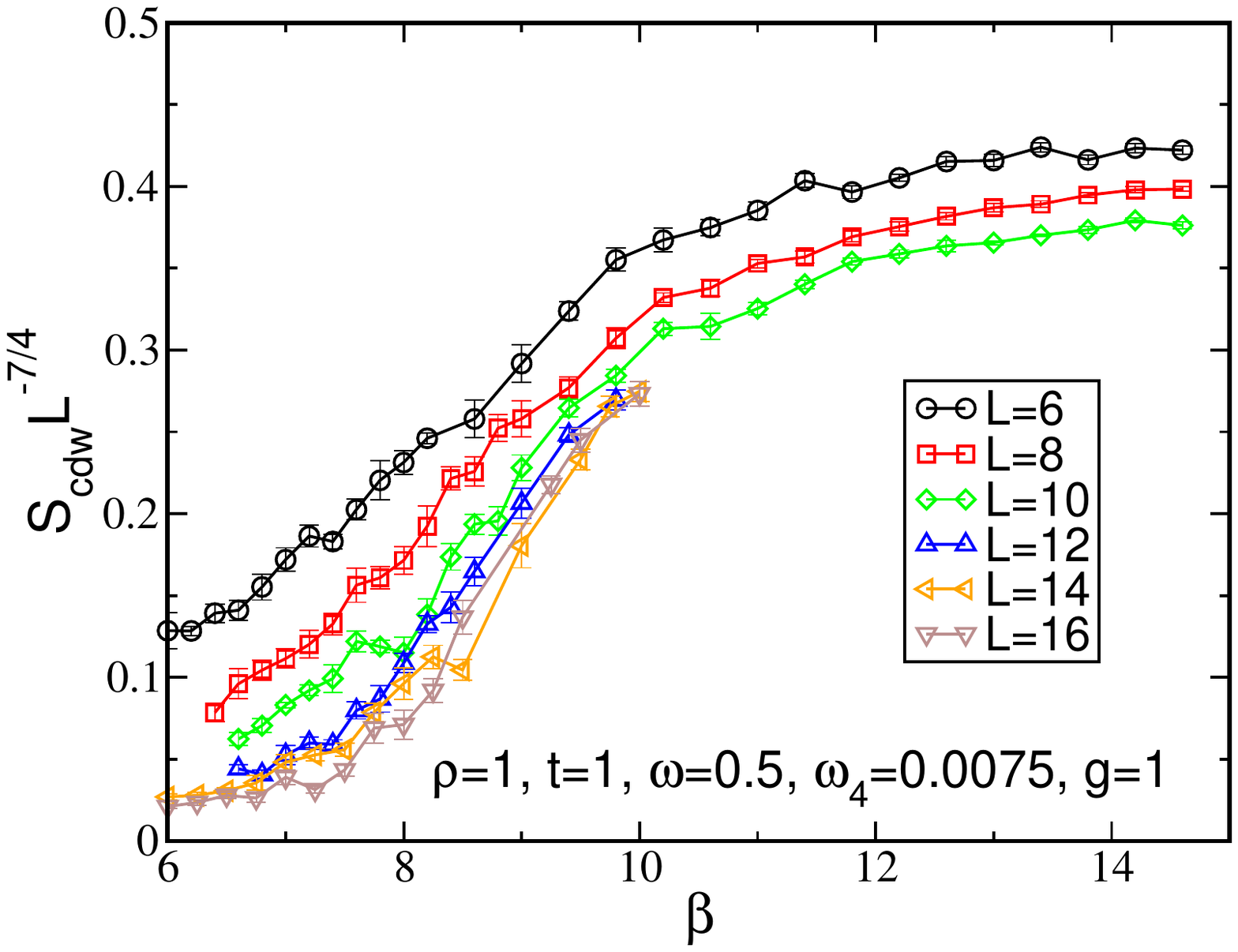}}
\caption{(Color online). For $\omega_4=0.0075$, $\omega=0.5$ and $L=6$
  to $L=16$, rescaled data for $S_{\rm cdw}$ do not cross each other,
  precluding the use of finite size scaling. This is probably due to
  larger finite size scaling corrections.
\label{fig:scalbad}}
\end{figure}

For values of $\omega_4$ larger than $\omega_4=0.005$ we have not been
able, in the range of temperatures and sizes that we could simulate,
to find cases where rescaled structure factor curves obtained for
different sizes would cross each other. Finite size analysis
(Eq. \ref{eq:scaling}) predicts that, for large systems, $S_{\rm
  cdw}\cdot L^{-7/4}$ should take a unique value $\tilde S(0)$ at
$T_c$.  For $\omega=0.5$ and $\omega_4=0.0075$, we studied system's
sizes up to $L=16$ for $\beta \le 10$ (see Fig.~\ref{fig:scalbad})
but, even for these relatively large systems, we could not find a
crossing point for the curves and could then not apply a finite size
scaling analysis to find the critical temperature.

\end{document}